\documentstyle [preprint,aps,epsf]{revtex}
\begin{document}
\title{Shot noise in coupled dots and the ``fractional charges''}
\author{Brahim Elattari$^{1}$ and S.A. Gurvitz $^2$\\
$^1$Universit\"{a}t Essen, Fachbereich 7, 45117 Essen, Germany\\
$^2$Weizmann Institute of Science, Department of Particle 
    Physics\\ 76100 Rehovot, Israel} 
\vspace{18pt}
\maketitle

\begin{abstract}
We consider the problem of shot noise in resonant tunneling  
through double quantum dots in the case of interacting particles. Using a 
many-body quantum mechanical description we evaluate the energy
dependent transmission probability, the total average current and the shot
noise spectrum. Our results show that the obtained reduction of the noise
spectrum, due to Coulomb interaction, can be interpret in terms of
non--interacting particles with fractional charge like behavior.    
\end{abstract}
 
\hspace{1.5 cm}  
\vspace{18pt}
 
The notion of quasi--particles of fractional charge has been 
introduced for almost two decades to explain  
the Fractional Quantum Hall (FQH) effect\cite{laughlin}. 
Yet, despite intensive efforts, the nature of these quasi--particles 
is not completely understood. An important progress, however, has been made 
in this direction with 
experiments on quantum shot-noise \cite{rez,sami,grif} leading to direct
measurement of the quasi--particle fractional charge. In fact, for 
non-interacting  particles of charge $q$ the zero frequency spectral 
density at zero temperature is given by\cite{les} 
 
\begin{equation}
S(0)=2qI(1-t)\ ,
\label{a0}
\end{equation}
where   
$I$ is the current and $t$ is the transmission coefficient through 
the device. In the FQH regime $q$ is given by the quasi--particle charge, 
$e^*$. 

For a proper understanding of the fractional charge, which appears in the 
shot noise, it is necessary to investigate a possible
modification of Eq.~(\ref{a0}) due to electron-electron interaction. 
Since the exact treatment of Coulomb interaction in FQH is a very complicated
problem, it would be interesting to investigate the shot noise in different 
transport processes, where the Coulomb interaction can be treated. 
These processes can be found, for example, in resonant tunneling  
structures, which are important not only from a fundamental,  
but also from an applied point of view. For instance, in was demonstrated 
that shot noise in the resonant tunneling diode, biased in the negative
differential resistance region may be described in terms of independent 
quasi-particles of charge up to $6.6q$\cite{inna1}. 

In this Letter we consider resonant transport through a  
two-level system, represented by two coupled dots, as shown 
schematically in Fig.~1, using a microscopic many-body description. 
The strong  Coulomb inter-dot 
repulsion prevents two electrons to occupy simultaneously the system. 
Hence an electron, entering this system turns to a linear 
superposition of the states of the two dots. The main question which 
we are going to consider is how the Coulomb interaction, which generates 
a partial occupation of the states of the system, 
affects the shot noise given by Eq.~(\ref{a0}). 
To answer of this question we apply our new method for a determination of  
the transmission $t$ in a case of interacting electrons.
We show that for
symmetric dots the effect of the Coulomb interaction
leads to a simple modification of Eq.~(\ref{a0}), in which the penetration
coefficient $t$ is replaced by $kt$, where $k$ is some fractional or integer factor,
depending on a particular system of dots.
Comparing with Eq.~(\ref{a0}) for non-interacting case we find that  
our result can be described by this equation, but assuming a fractional
charge like behavior, similar to that observed in the FQH regime. 
Even though this system is very different from that of FQH, 
the similarity in the behavior of
shot-noise in different transport processes would be useful to
understand the nature of the fractional charge quasi-particles\cite{jack}. 

Let us consider resonant tunneling through
two levels of coupled quantum dots connected to two separate reservoirs 
of very dense states, as  shown in Fig.~1. The
reservoirs are taken at zero temperature and are filled up to the Fermi
energy levels $\mu_L$ and $\mu_R$, respectively, with $\mu_L > \mu_R$. As a
result the current flows from left to the right reservoirs.
\vskip1cm
\begin{minipage}{13cm}
\begin{center}
\leavevmode
\epsfxsize=10cm
\epsffile{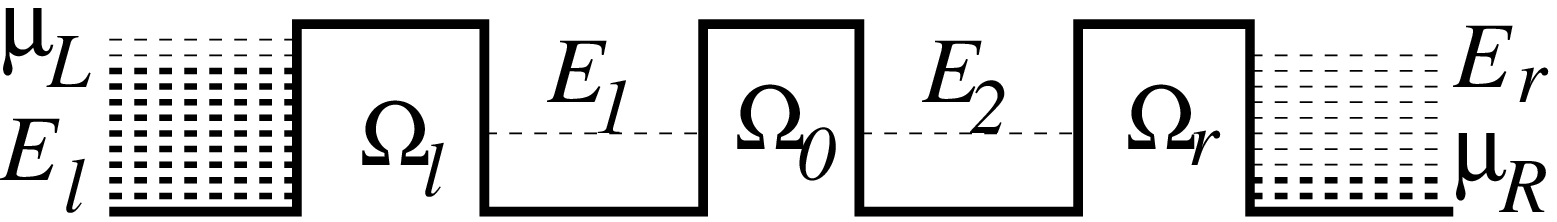}
\end{center}
{\begin{small}
{\bf Fig.~1:}  Resonant tunneling through coupled dots. 
\end{small}}
\end{minipage} \\ \\
 
The dynamics of the system is described by the following tunneling Hamiltonian,
where for simplicity we have omitted the spin variables 
\begin{eqnarray} 
H &=& \sum_l E_{l}a^{\dagger}_{l}a_{l} +              
E_1 a_1^{\dagger}a_{1} + E_2 a_2^{\dagger}a_{2}+                
\sum_r E_{r}a^{\dagger}_{r}a_{r}
+\sum_l \Omega_{l}(a^{\dagger}_{l}a_1
+a^{\dagger}_{1}a_{l})\nonumber\\            
&&+ \sum_r \Omega_{r}(a^{\dagger}_{r}a_2 +              
a^{\dagger}_{2}a_{r})           
+\Omega_0(a_1^{\dagger}a_{2}+a_2^{\dagger}a_{1})               
+Ua_1^{\dagger}a_1a_2^{\dagger}a_2 \ . 
\label{b1} 
\end{eqnarray} 
The operators $a_i^{\dagger}$ ($a_i$) create (annihilate) an
electron in the corresponding state $i$. The $\Omega_l$ and  $\Omega_r$   are 
the hopping amplitudes between the states $E_l$, $E_1$ and $E_r$,$E_2$, 
respectively. These amplitudes are found to be directly related to the
tunneling rate of the electrons out of the quantum dots,
$\Gamma_{L,R}=2\pi\Omega^2_{L,R}\rho_{L,R}$, where 
$\Omega_{L,R}=\Omega_{l,r}(E_{1,2})$ and $\rho_{L}$ ($\rho_{R}$) is the 
density of states in the left (right) reservoir. 
The $\Omega_0$ is the hopping amplitude between the dots. 
The Coulomb interaction between
the two dots is given by the last  term in Eq.~(\ref{b1}).   
The intra-dot Coulomb repulsion is assumed to be very large, so that only
one electron may occupy a dot. 
 
It was demonstrated in Ref.\cite{gp} that in the case of large bias,
$\mu_L-\mu_R\gg\Gamma_{L,R}, \Omega_0$, the time dependent Schr\"odinger
equation $i|\dot\Psi(t)>=H|\Psi(t)$, where $|\Psi>$ is the many-body 
wave function of the above system, can be transformed to a set of rate
equations for the corresponding density matrix, 
$\sigma^{(n)}_{ij}(t)$. 
This density matrix, gives the probability of finding $n$ electrons in the 
collector by time $t$, where the indices 
$i,j=\{0,1,2,3\}$ denote the different possibilities of occupation of the 
states of the dots, namely: 
$\sigma_{00}^{(n)}$ is the probability  of finding both dots unoccupied, 
$\sigma_{11}^{(n)}$, $\sigma_{22}^{(n)}$, $\sigma_{33}^{(n)}$ are the 
probabilities of 
finding the first, the second, and both of the dots occupied, 
respectively;  $\sigma_{12}^{(n)}(t)=\sigma_{21}^{* (n)}(t)$ denote 
the off-diagonal density-matrix elements (the ``coherencies''). 

Consider first the non-interacting case, $U=0$ in Eq.~(\ref{b1}). 
Using our approach we derive the 
following Bloch-type rate equations for $\sigma^{(n)}_{ij}(t)$,
which have a simple interpretation in term of the ``loss'' and
``gain'' contributions\cite{gp}
\begin{mathletters} 
\label{b2} 
\begin{eqnarray}
\dot\sigma_{00}^{(n)} & = & -\Gamma_L\sigma_{00}^{(n)}
+\Gamma_R\sigma_{22}^{(n-1)}\;, \label{b2a}\\
\dot\sigma_{11}^{(n)} & = & \Gamma_L\sigma_{00}^{(n)}
+\Gamma_R\sigma_{33}^{(n-1)}+i\Omega_0(\sigma_{12}^{(n)}-\sigma_{21}^{(n)})\;,
\label{b2b}\\
\dot\sigma_{22}^{(n)} & = & -\Gamma\sigma_{22}^{(n)}
+i\Omega_0(\sigma_{21}^{(n)}-\sigma_{12}^{(n)})\;,
\label{b2c}\\
\dot\sigma_{33}^{(n)} & = & -\Gamma_R\sigma_{33}^{(n)}
+\Gamma_L\sigma_{22}^{(n)}\;,
\label{b2d}\\
\dot\sigma_{12}^{(n)} & = & i\epsilon\sigma_{12}^{(n)}+
i\Omega_0(\sigma_{11}^{(n)}-\sigma_{22}^{(n)})
-\Gamma\sigma_{12}^{(n)}\ ,
\label{b2e}
\end{eqnarray}
\end{mathletters}
where $\epsilon =E_2-E_1$ and $\Gamma =(\Gamma_L+\Gamma_R)/2$. 

The stationary current flowing in the system is then given by: 
\begin{equation}
\left. I=e{d\over dt}\sum_nnP_n(t)\right |_{t\to\infty},
\label{b3}
\end{equation}
where $P_n(t)=\sum_{i=0}^{i=3}\sigma^{(n)}_{ii}(t)$
is the total probability of finding $n$ electrons in the 
collector by time $t$. The zero frequency component
of the noise spectrum is defined by $P_n(t)$ as well\cite{chen}:
\begin{equation}
S(0)=2e^2\left.{d\over dt}\left [\sum_nn^2P_n(t)-(\sum_nnP_n(t))^2\right ]
\right |_{t\to\infty}
\label{b5}
\end{equation}

Solving Eqs.~(\ref{b2}) in the limit of
$t\to\infty$ and using Eqs.~(\ref{b3}), (\ref{b5}) one finds 
the following expressions for the averaged current
and the shot noise power, respectively 
\begin{eqnarray}
I&=&e{2\Gamma_L\Gamma_R\Gamma\Omega_0^2\over
(4\Omega_0^2+\Gamma_L\Gamma_R)\Gamma^2
+\epsilon^2\Gamma_L\Gamma_R}\ ,
\label{b4}\\\noalign{\vskip7pt}
S(0)&=&2eI\bigg (1-\Gamma_L\Gamma_R\Omega_0^2
{\epsilon^2(\Gamma_L^3+\Gamma_R^3)
+2\Gamma^3(4\Gamma^2+\Gamma_L\Gamma_R+4\Omega_0^2)\over
\Gamma [(4\Omega_0^2+\Gamma_L\Gamma_R)\Gamma^2
+\epsilon^2\Gamma_L\Gamma_R]^2}\bigg ).
\label{b7} 
\end{eqnarray}

Since Eqs.~(\ref{b4}) and (\ref{b7}) correspond to
the non-interacting case, $U=0$, one can show that the same 
expressions for $I$ and $S(0)$ can be obtained in an alternative way 
using the Landauer formula for $I$ 
and the Khlus-Lesovik formula for $S(0)$,  
\begin{eqnarray}
I&=&e\int {dE\over 2\pi} t(E)\ ,
\label{b88}\\
S(0)&=&2e^2\int
{dE\over 2\pi}t(E)(1-t(E))\ .
\label{b8}
\end{eqnarray}
Here $t(E)$ is the transmission probability through the
sample,
given by the Breit-Wigner formula for coupled 
resonances\cite{sum}
\begin{equation}
t(E)={\Gamma_L\Gamma_R\Omega_0^2\over \left |\left
(E-E_1+i{\Gamma_L\over 2}\right ) \left (E-E_2+i{\Gamma_R\over 2}\right
)-\Omega_0^2\right |^2} . 
\label{b9}
\end{equation}

In the case of small coupling with the reservoirs,
$\Gamma_{L,R}\ll\Omega_0$ and $\epsilon \ll \Omega_0$ ,  
the transmission $t(E)$
shows two pronounced peaks at $E=E_\pm$, corresponding to the two 
eigenstates of the double-dot system. One easily finds that in 
this limit Eq.~(\ref{b7}) for the shot noise becomes
\begin{equation} 
S(0)=2eI( 1-\bar t/2) ,
\label{nb}
\end{equation}
where $\bar t=t(E_\pm )=4\Gamma_L\Gamma_R/(\Gamma_L+\Gamma_R)^2$ is the
maximum  value of the transmission probability. Hence, Eq.~(\ref{nb}) 
predicts a maximum suppression of one half of the Poissonian 
noise (for $\Gamma_L=\Gamma_R$), in agreement with 
the known results\cite{ian} for independent particle models. 

Let us apply our approach to the case where $U\to\infty$ in Eq.~(\ref{b1}).
This corresponds to strong interdot Coulomb repulsion, which   
prevents simultaneous occupation of the two quantum dots. As a
result the associated rate equations become \cite{gp}:
\begin{mathletters} 
\label{b10} 
\begin{eqnarray}
\dot\sigma_{00}^{(n)} & = & -\Gamma_L\sigma_{00}^{(n)}
+\Gamma_R\sigma_{22}^{(n-1)}\;, \label{b10a}\\ 
\dot\sigma_{11}^{(n)} & = &
\Gamma_L\sigma_{00}^{(n)} +i\Omega_0(\sigma_{12}^{(n)}-\sigma_{21}^{(n)})\;,
\label{b10b}\\
\dot\sigma_{22}^{(n)} & = & -\Gamma_R\sigma_{22}^{(n)}
+i\Omega_0(\sigma_{21}^{(n)}-\sigma_{12}^{(n)})\;,
\label{b10c}\\
\dot\sigma_{12}^{(n)} & = & i\epsilon\sigma_{12}^{(n)}+
i\Omega_0(\sigma_{11}^{(n)}-\sigma_{22}^{(n)})
-\frac{1}{2}\Gamma_R\sigma_{12}^{(n)}\;.
\label{b10d}
\end{eqnarray}
\end{mathletters}

Solving Eqs.~(\ref{b10}) in the limit of $t\to\infty$ and using 
Eq.~(\ref{b3}) we find  the following expression for the total current,
in the case of Coulomb interdot blockade\cite{gp,naz}:
\begin{equation}
I_c=e{4\Gamma_L\Gamma_R\Omega_0^2\over
4\Omega_0^2(2\Gamma_L+\Gamma_R)+\Gamma_L\Gamma_R^2
+4\epsilon^2\Gamma_L}.
\label{b11}
\end{equation}
The corresponding shot noise spectrum, $S_c(0)$, is obtained 
in an analogous way
by Eq.~(\ref{b5}), leading to  
\begin{eqnarray}
S_c(0)=2eI_c\bigg (1-8\Gamma_L\Omega_0^2
{4\epsilon^2(\Gamma_R-\Gamma_L)+
\Gamma_R(3\Gamma_L\Gamma_R+\Gamma_R^2+8\Omega_0^2)\over
[4\Omega_0^2(2\Gamma_L+\Gamma_R)+\Gamma_L\Gamma_R^2
+4\epsilon^2\Gamma_L]^2}\bigg )\ .
\label{b12}
\end{eqnarray}

Now, it is interesting to compare $S_c(0)$, given by Eq.~(\ref{b12}), with
that obtained from the Khlus-Lesovik formula Eq.~(\ref{b8}). The crucial 
problem, however, is an evaluation of the transmission
coefficient for interacting electrons, $t_c(E)$, given by the ratio
between the conductance and the quantum conductance\cite{rez}. Thus,  
$t_c(E)={\cal I}(E)/e$, where ${\cal I}(E)$ is  
the current density ($I_c=\int {\cal I}(E)dE/2\pi$), 
represented by the operator $ie[H,a_r^\dagger a_r]=
ie\Omega_r(a^\dagger_2a_r-a^\dagger_ra_2)$,
where $E=E_r$ is the corresponding energy level in the collector.
One finds
\begin{equation}
{\cal I}(E_r)=ie\Omega_r [\sigma_{r2}(t)-\sigma_{2r}(t)]|_{t\to\infty}, 
\label{bbb12}
\end{equation}
where $\sigma_{r2}(t)=\langle \Psi (t)|a^\dagger_2a_r|\Psi (t)\rangle$ 
are the off-diagonal density-matrix elements describing an electron in the 
linear superposition between the states $E_2$ and $E_r$.  
Such terms cannot be determined by the rate equations (\ref{b10}),
where all the reservoir states $E_{l,r}$ were traced out. 
Yet, it was recently found that our method can be generalized, by avoiding  
any tracing of the collector states, $E_r$\cite{eg}. Finally we arrive 
to extended quantum rate equations, which determine the   
off-diagonal terms, $\sigma_{r2}(t)=\sigma^*_{2r}(t)$ and 
$\sigma_{r1}(t)=\sigma^*_{1r}(t)$:
\begin{mathletters}
\label{b14}
\begin{eqnarray}
\dot{\sigma}_{r1}&=&i(E_1-E_r)\sigma_{r1}+i\Omega_0\sigma_{r2}
-i\Omega_r\sigma_{21}-{\Gamma_L\over 2}\sigma_{r1}
\label{b14a}\\
\dot{\sigma}_{r2}&=&i(E_2-E_r)\sigma_{r2}
+i\Omega_0\sigma_{r1}-i\Omega_r\sigma_{22}
-\Gamma\sigma_{r2}\ ,
\label{b14b}
\end{eqnarray}
\end{mathletters}
where $\sigma_{21}=\sum_n\sigma_{21}^{(n)}$ and
$\sigma_{22}=\sum_n\sigma_{22}^{(n)}$ are given by Eqs.~(\ref{b10}).
The detailed derivation of these equations will be given 
elsewhere. Here we only mention that these equations resembles 
the Bloch equation, with some modifications, since the states $E_r$
belong to continuum\cite{eg}.

Solving Eqs.~(\ref{b10}), (\ref{b14}) and using Eq.~(\ref{bbb12})
we obtain for the current density the
following  result 
\begin{equation}
{\cal I}(E)={\Gamma_L[(E-E_2)^2+\Gamma^2]
+2\Omega_0^2\Gamma\over \left |
\left (E-E_1+i{\Gamma_L\over 2}\right )
\left (E-E_2+i\Gamma\right )-\Omega_0^2\right |^2}I_c , 
\label{b15}
\end{equation}
where $I_c$ is the total current, given by Eq.~(\ref{b11}).
Similar to the previous case of non-interacting electrons, Eq.~(\ref{b9}), 
the transmission coefficient $t_c(E)={\cal I}(E)/e$ shows two
peaks in the limit of weak coupling of the quantum dots with 
the reservoirs, $\Gamma_{L,R}\ll \Omega_0$, and $\epsilon\ll\Omega_0$. 
Each of these peaks is well reproduced by a Lorentzian given by 
\begin{equation}
t_c(E)={\Gamma_L\Gamma_R/4
\over (E-E_{\pm})^2
+\left ({2 \Gamma_L +\Gamma_R \over 4}\right )^2}\ ,
\label{bb15}
\end{equation}
as one can show directly from Eq.~(\ref{b15}) in this limit.
However, in the limit of strong coupling or $\epsilon\gg\Omega_0$,  
the transmission coefficient is rather flat. This is demonstrated in Fig.~2,
which shows $t_c(E)$ for a symmetric system 
($\Gamma_L=\Gamma_R =\Gamma$, $E_1=E_2=0$). 
It is important to note that in contrast to non-interacting 
electrons, the height of the peaks is always smaller than $1$, 
since the interdot interaction does not
allow simultaneous occupation of both quantum dots. 
\vskip1cm
\begin{minipage}{13cm}
\begin{center}
\leavevmode
\epsfxsize=10cm
\epsffile{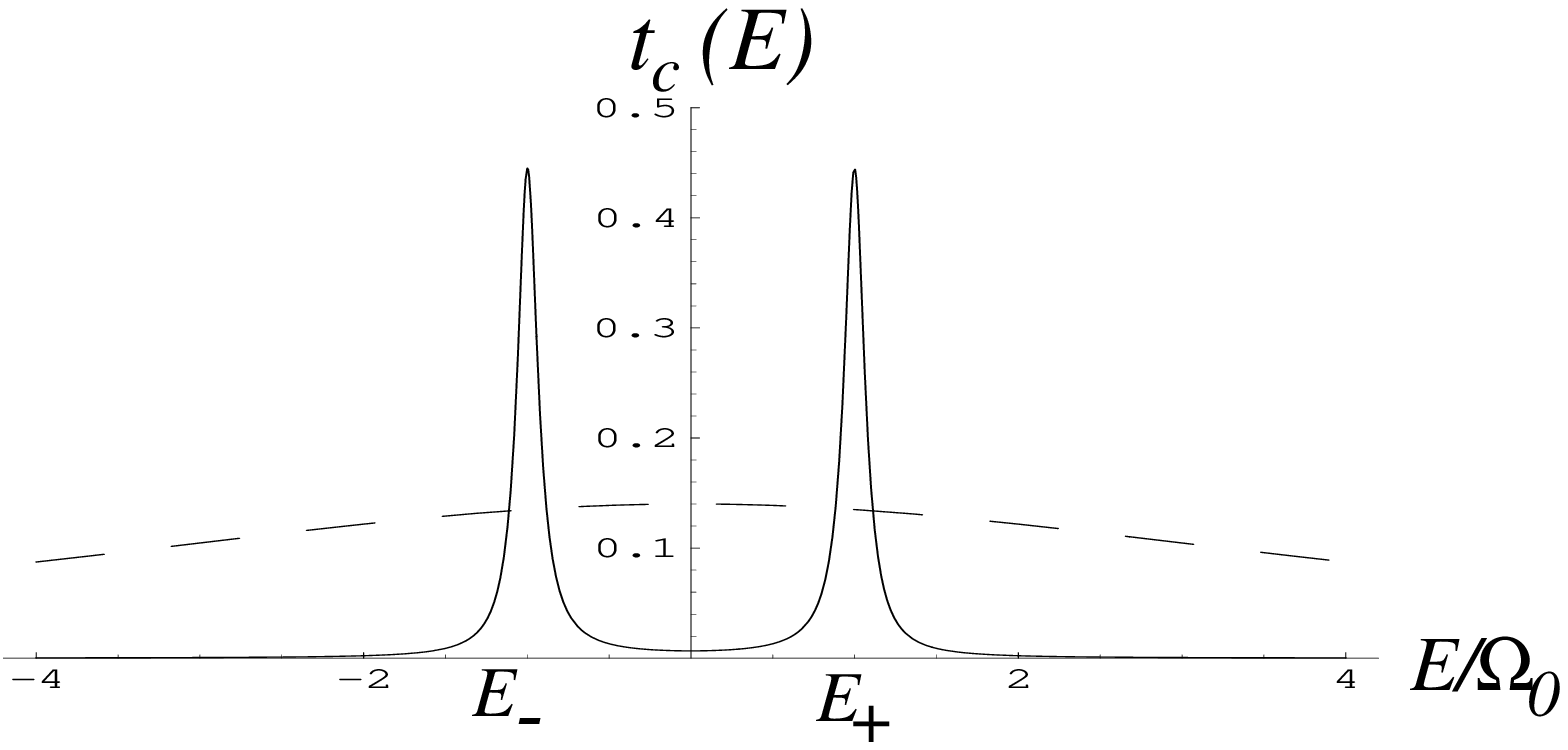}
\end{center}
{\begin{small}
{\bf Fig.~2:} Transmission $t_c(E)$  for a symmetric 
coupled-dot in the case of interacting electrons ($U\to\infty$). 
The solid line corresponds to $\Gamma =0.1\Omega_0$,
whereas the dashed line corresponds to $\Gamma =10\Omega_0$.
\end{small}}
\end{minipage} \\ \\

Substituting $t_c={\cal I}/e$ given by Eq.~(\ref{b15}) into Eq.~(\ref{b8})
and comparing the result with $S_c(0)$, Eq.~(\ref{b12}) for
$\epsilon\ll \Omega_0$, we find  
a further suppression of the shot noise due to Coulomb interaction 
with respect to that obtained from the Khlus-Lesovik formula. 
This is demonstrated in Fig.~3, where we plot the shot-noise as  
a function of $\epsilon$. With an increase   
of $\epsilon$, however, the opposite effect takes place, and the 
non-interacting formula Eq.~(\ref{b8}) shows suppression of the 
shot noise with respect to the interacting case. 
For $\epsilon\to \infty$ both Eq.~(\ref{b8}) 
and Eq.~(\ref{b12}) reach the same Poissonian limit. 
\vskip1cm
\begin{minipage}{13cm}
\begin{center}
\leavevmode
\epsfxsize=10cm
\epsffile{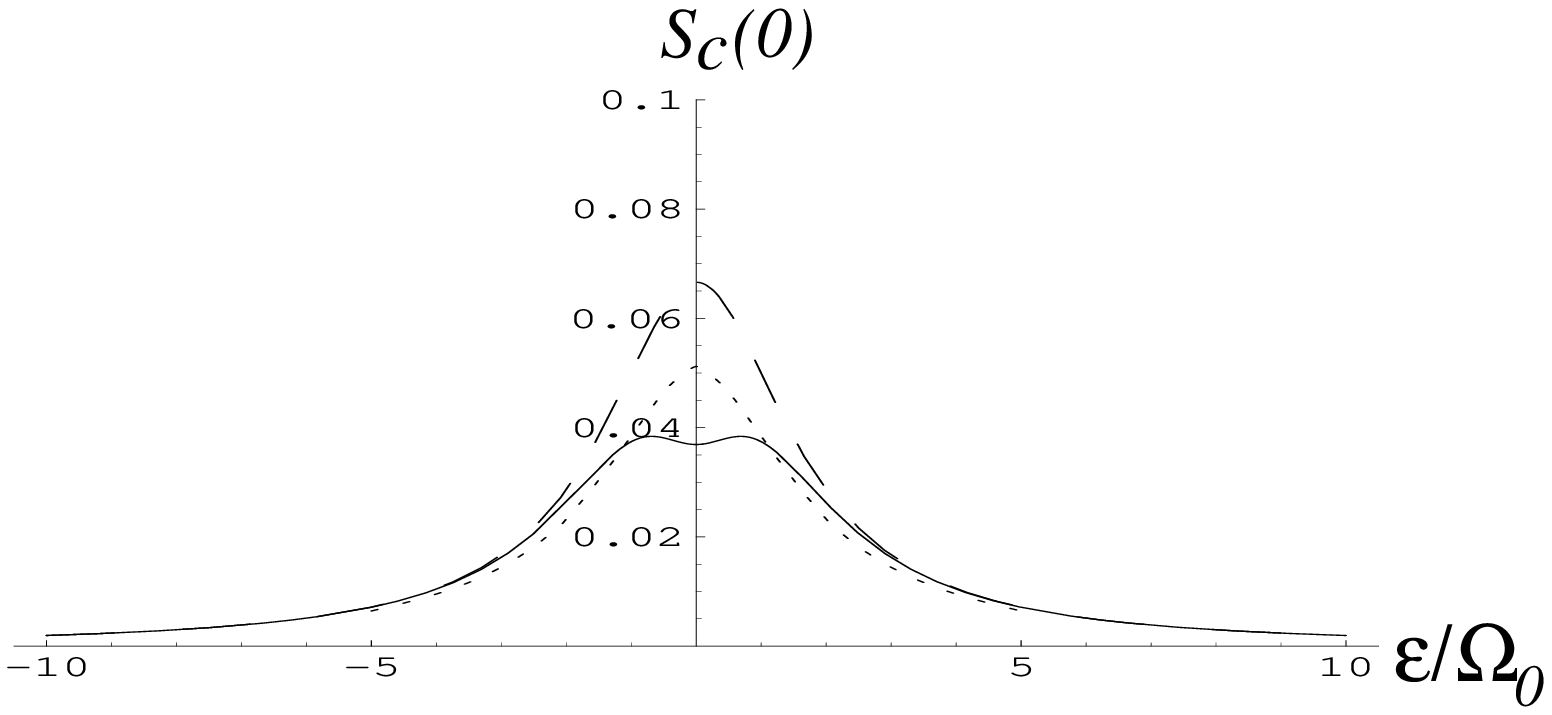}
\end{center}
{\begin{small}
{\bf Fig.~3:} The shot-noise power as a function of 
$\epsilon =E_1-E_2$ for $U\to\infty$ and 
$\Gamma_L=\Gamma_R=0.1\Omega_0$, 
The solid line corresponds to Eq~(\ref{b12}), the dashed line
is the Schottky noise, and the dotted line is given by 
Eqs.~(\ref{b8}), (\ref{b15}).
\end{small}}
\end{minipage} \\ \\

The additional reduction of the shot noise due to Coulomb repulsion,  
with respect to the non-interacting case, can be 
seen from the simple analytical results obtained in the limit of weak coupling  
with the reservoirs, $\Gamma_{L,R}\ll \Omega_0$
and $\epsilon\ll\Omega_0$. One finds from Eq.~(\ref{b12})
\begin{equation} 
S_c(0)=2eI_c( 1-\bar t_c) ,
\label{ni}
\end{equation}
where $\bar t_c$ is the peak value of the transmission probability
\begin{equation}
\bar t_c=t(E_\pm )={4\Gamma_L/\Gamma_R\over (2\Gamma_L/\Gamma_R+1)^2}
\label{bb16}
\end{equation}  
On the other hand the Khlus-Lesovik formula (\ref{b8}) and Eq.~(\ref{b15})
give in the same limit  
\begin{equation} 
S_0(0)=2eI_c( 1-\bar t_c/2) ,
\label{ni1}
\end{equation}
 Comparing Eqs.~(\ref{ni})
and (\ref{ni1}), it follows that an additional reduction of the shot-noise due 
to Coulomb
blockade can by effectively accounted for by a factor $k$ in the second
term of the Khlus-Lesovik formula, i.e. by replacing $(1-t)$ in Eq.~(\ref{b8})
by $(1-kt)$. We suggest that this result would be valid in the general
case of different configurations and number of quantum dots. Since 
this reduction of the shot noise is associated to  
a partial occupation of the available states, we can expect that 
$k$ is always given by an integer or fraction.

Now we are going to investigate whether one could interpret the 
result given by Eq.~(\ref{b12}) in terms of non-interacting particles
with a fractional charge $e^*$. To answer this question, let as compare 
Eq.~(\ref{b12}) with Eq.~(\ref{b8}), where in the latter the electron charge 
$e$ is replaced by the quasi-particle charge $e^*$. Respectively, the
transmission $t(E)$ is replaced by ${\cal I}(E)/e^*$
with the same quasi-particle charge $e^*$. This
allows us to determine the transmission via the current, as in
the measurements\cite{rez,sami,grif}. Thus Eq.~(\ref{b8}) can be rewritten as
\begin{equation} 
S_0(0)=2e^*\int {\cal I}(E)\left [1-{{\cal I}(E)\over e^*}\right ]
{dE\over 2\pi}    
\label{bb12} 
\end{equation}

Consider again the limit of $\Gamma_{L,R},\epsilon\ll\Omega_0$.
Then using Eg.~(\ref{bb15}), we find from Eq.~(\ref{bb12}) that $S_0(0)$ 
is given by Eq.~(\ref{ni1}) with $e\to e^*$ and $\bar t_c\to (e/e^*)\bar t_c$.
To determine the value of the charge $e^*$, we require this expression 
to be identical with the one given by Eq.~(\ref{ni}). We then obtain the
following expression for $e^*/e$
as a function of the transmission 
$t^*={\cal I}(E_\pm )/e^*=(e/e^*)\bar t_c$,
\begin{equation}
\frac{e^*}{e}=\frac{1}{1+pt^*} ,
\label{n4}
\end{equation}
where $p=1/2$. It follows from Eq.~(\ref{bb16}) that 
the value of $t^*$ depends on  
the ratio $\Gamma_L/\Gamma_R$, but cannot exceed $t^*_{max}=2/3$. 
This corresponds to the fractional charge $e^*/e=3/4$.
In general, $t^*_{max}$, as well as the factor $p$ in
Eq.~(\ref{n4}) depend on the particular geometry (whether the dots are in
sequel or in parallel) and on the number of coupled dots.
We thus would assume that Eq.~(\ref{n4}) holds in the general case, 
where only the coefficient $p$ is different. Since the latter is related to
partial occupation of the states, we expect it to be either an integer or a
fractional number. 

It is very interesting that Eq.~(\ref{n4}) for $p=2$ is rather close to the
experimental results for the measurement of the fractional charge\cite{grif}
in the FQH at filling factor $\nu =1/3$, as shown in Fig.~4.
Although the processes are very different, a resemblance between Eq.~(\ref{n4})
and the data might indicate the importance of partial occupation of
available states in FQH effect.

In conclusion, we have given a microscopic description of the shot noise in
resonant structures in the presence of Coulomb interaction. Our results show 
that for symmetric dots Coulomb interaction decreases the noise in comparison to the
non-interacting case and the obtained shot noise can be interpret in terms of
non-interacting particles with fractional charge due to a partial occupation
of the quantum dots states.  

\vskip1cm
\begin{minipage}{13cm}
\begin{center}
\leavevmode
\epsfxsize=10cm
\epsffile{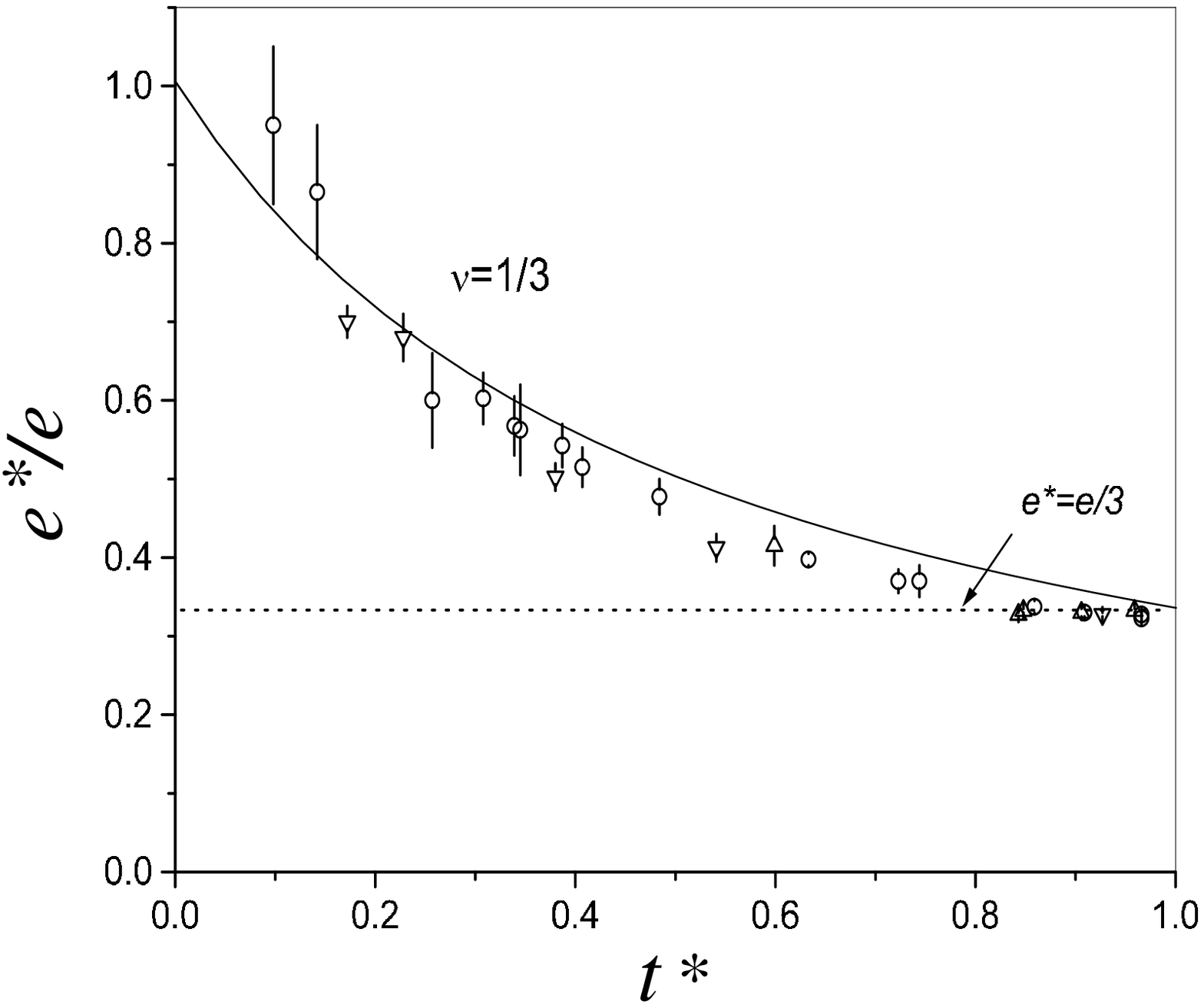}
\end{center}
{\begin{small}
{\bf Fig.~4:} The quasi-particle charge in FQH as a function of transmission.
The solid line corresponds to Eq.~(\ref{n4}) for $p=2$.
\end{small}}
\end{minipage} \\ \\

BE knowledges fruitful discussions with G. Hackenbroich and would like 
to thank the Albert Einstein Minerva Center for Theoretical 
Physics for the financial support.

\end{document}